\RequirePackage{fix-cm}
\documentclass[twocolumn,epjc3]{svjour3}  
\smartqed  
\RequirePackage{graphicx}
\RequirePackage{latexsym}
\RequirePackage[numbers,sort&compress]{natbib}
\RequirePackage[colorlinks,citecolor=blue,urlcolor=blue,linkcolor=blue]{hyperref}
\usepackage{amssymb,amsfonts}
\journalname{Eur. Phys. J. C}
\begin{document}

\title{On the Schwarzschild-Anti-de Sitter black hole with an f(R) global monopole
}


\titlerunning{SAdS black hole with with an f(R) global monopole}        

\author{H. S. Vieira\thanksref{e1,e2,addr1}}

\thankstext{e1}{e-mail: horacio.santana.vieira@hotmail.com}
\thankstext{e2}{e-mail: horacio.santana-vieira@tat.uni-tuebingen.de}


\institute{Theoretical Astrophysics, Institute for Astronomy and Astrophysics, University of T\"{u}bingen, 72076 T\"{u}bingen, Germany \label{addr1}}

\date{Received: \today / Accepted: date}

\maketitle

\begin{abstract}
In this work, we follow the recently revisited f(R) theory of gravity for studying the interaction between quantum scalar particles and the gravitational field of a generalized black hole with an f(R) global monopole. This background has a term playing the role of an effective cosmological constant, which permits us to call it as Schwarzschild-Anti-de Sitter (SAdS) black hole with an f(R) global monopole. We examine the separability of the Klein-Gordon equation with a non-minimal coupling and then we discuss both the massless and massive cases for a conformal coupling. We investigate some physical phenomena related to the asymptotic behavior of the radial function, namely, the black hole radiation, the quasibound states, and the wave eigenfunctions.
\keywords{modified gravity \and Klein-Gordon equation \and Heun function \and Hawking radiation \and resonant frequency \and eigenfunction}
\end{abstract}

%
%
\section{Introducing the Schwarzschild-Anti-de Sitter black hole with an f(R) global monopole}\label{SAdS}
In a recent paper, Caram\^{e}s \textit{et. al.} \cite{EurPhysJC.77.496} obtained a class of exact solutions for the modified field equations in the presence of a global monopole for regions outside its core, which generalize some previous results in the f(R) theory of gravity. In this section, we will give a brief review about their results and then setup the background in which we want to investigate the behavior of quantum scalar fields following the Vieira-Bezerra-Kokkotas method \cite{AnnPhys.373.28,PhysRevD.104.024035}.

This paper has three broad goals. First, to obtain a metric which describes the Schwarzschild-Anti-de Sitter black hole with an f(R) global monopole. Second, to discuss the separability of the Klein-Gordon equation in the background under consideration. Third, to compute the Hawking radiation, the quasibound states, and the wave eigenfunctions.

The action of the f(R) theory of gravity is given by
\begin{equation}
S=\frac{1}{2\kappa} \int d^{4}x \sqrt{-g} [f(R)+\mathcal{L}],
\label{eq:action_f(R)}
\end{equation}
where $g$ is the determinant of a $4 \times 4$ matrix constructed from the metric tensor, $\mathcal{L}$ is the Lagrangian density, and $\kappa=8\pi$. Note that we are adopting the natural units, namely, $G \equiv c \equiv \hbar \equiv 1$. Now, this action is extremized with respect to the metric tensor, which leads to the following field equations
\begin{eqnarray}
\kappa T_{\mu\nu} & = & F(R)R_{\mu\nu}-\frac{1}{2}f(R)g_{\mu\nu}-\nabla_{\mu}\nabla_{\nu}[F(R)]\nonumber\\
&& +\Box[F(R)]g_{\mu\nu},
\label{eq:field_equations_f(R)}
\end{eqnarray}
where $F(R)=df(R)/dR$. On the other hand, the global monopole spacetime model is described by the following Lagrangian density
\begin{equation}
\mathcal{L}=\frac{1}{2}\partial_{\mu}\phi^{a}\partial^{\mu}\phi^{a}-\frac{1}{4}\lambda(\phi^{a}\phi^{a}-\eta^{2})^{2},
\label{eq:Lagrangian_density_f(R)}
\end{equation}
where $\lambda$ is a positive coupling constant, $\eta$ is the energy scale at which the symmetry is broken, and the Higgs field $\phi^{a}$ is given by an isotriplet of scalar fields, whose form corresponds to the well-known hedgehog \textit{Ansatz}. Then, a spherically symmetric line element, which describes the spacetime around a static source, can be written, in general, as
\begin{equation}
ds^{2}=-B(r)\ dt^{2}+A(r)\ dr^{2}+r^{2}\ d\theta^{2}+r^{2}\sin^{2}\theta\ d\phi^{2},
\label{eq:spherically_symmetric_f(R)}
\end{equation}
where $A(r)$ and $B(r)$ are functions to be determined; they are related by $Y(r) \equiv A(r)B(r)$. In this model, the energy-momentum tensor has a very simple form. It is given by
\begin{equation}
T_{\mu}^{\nu} \approx \mbox{diag} \biggl(\frac{\eta^{2}}{r^{2}},\frac{\eta^{2}}{r^{2}},0,0\biggr).
\label{eq:energy-momentum tensor_f(R)}
\end{equation}
Now, an alternative parametrization for $F(R)$ is a\-dopt\-ed, namely, $F(R(r))=\mathcal{F}(r)=1+\psi(r)$, where $\psi(r)=\psi_{0}r$ is a function encoding the deviation from the Einstein's general relativity. From this parametrization, we get
\begin{equation}
Y(r) = A(r) B(r) = Y_{0},
\label{eq:alternative_parametrization}
\end{equation}
where $Y_{0}$ is a constant. Thus, after some algebra, it can be found the following expression for the metric coefficient $B(r)$:
\begin{eqnarray}
B(r) & = & Y_{0}(1-8 \pi \eta^{2})+\biggl(\frac{\psi_{0}}{2}-\frac{1}{3r}\biggr)c_{1}\nonumber\\
&& -r\psi_{0}[Y_{0}(1-16 \pi \eta^{2})+\psi_{0}c_{1}]\nonumber\\
&& +\frac{r^{2}}{2}\biggl\{\psi_{0}^{2}Y_{0}(3-32 \pi \eta^{2})+2c_{2}\nonumber\\
&& +2\psi_{0}^{2}[Y_{0}(1-16 \pi \eta^{2})+\psi_{0}c_{1}]\ln\biggl(\psi_{0}+\frac{1}{r}\biggr)\biggr\},
\label{eq:B(r)_f(R)}
\end{eqnarray}
where $c_{1}$ and $c_{2}$ are constants (of integration) to be determined (or to be opportunely chosen). In fact, it is worth noticing that this solution is more general than that ones found in the literature, since it carries corrections that are absent in all other approaches (including the ones where the approximations $|\psi_{0}r| \ll 1$ and the weak field limit were taken into account). Thus, Caram\^{e}s \textit{et al.} found a set of black hole solutions, which are displayed in Table 1 of Ref. \cite{EurPhysJC.77.496}.

Now, we will take some useful approximations into account in order to establish a new particular black hole background. This will be possible due to the suitable choice of the constants $c_{1}$ and $c_{2}$, as follows. First of all, we set $Y_{0}=1$, which implies that $A(r)=[B(r)]^{-1}$. Then, by assuming a small correction on the Einstein's general relativity, we can keep just the linear terms in $\psi_{0}r$ by considering the constant $\psi_{0}$ very tiny, which means that $\psi_{0}^{2} \sim 0$. Furthermore, we can throw away all the crossing terms involving $\psi_{0}$ and $\eta^{2}$. Thus, we can choice $c_{1}=6M$ and $c_{2}=\tilde{\Lambda}/3$, which means that they are associated to the Newtonian potential and to the effects of an effective cosmological constant, respectively (for details, see Ref.~\cite{IntJTheorPhys.45.2357} and references therein). Finally, we get
\begin{equation}
B(r)=1-8 \pi \eta^{2}+3M\psi_{0}-\frac{2M}{r}-\psi_{0}r+\frac{\tilde{\Lambda}}{3}r^{2},
\label{eq:function_B(r)_SAdS}
\end{equation}
where $M$ represents the total mass centered at the origin of the system of coordinates, $\tilde{\Lambda}$ is an effective cosmological constant playing the same role as the standard cosmological constant ($3/\ell^{2}$) in the dynamics of the universe. Therefore, we have obtained a metric corresponding to the SAdS black hole with an f(R) global monopole. From now on, due to the choice of approximations described above, we will use the following values for the involved parameters: $\psi_{0}=0.02$, $8 \pi \eta^{2}\sim\eta^{2}=10^{-6}$, $\tilde{\Lambda}=0.12$ $(\ell_{\rm ISCO}=5)$, and $M=1$. In fact, these are the values expected within the Grand Unified Theories (GUT) for the potential appearance of topological defects in the early universe.
%
%

\begin{figure}
		\includegraphics[width=0.45\textwidth]{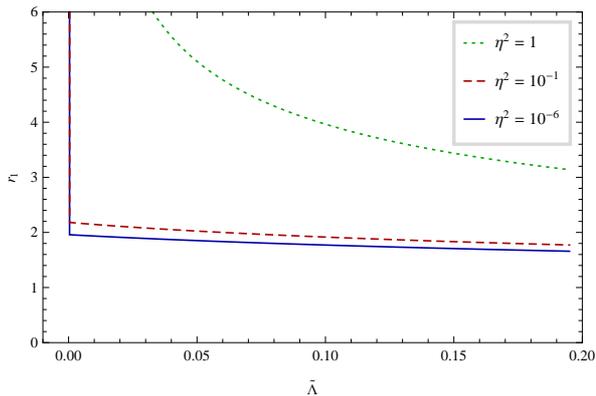}
	\caption{The exterior event horizon $r_{1}$ as a function of $\tilde{\Lambda}$.}
	\label{fig:Fig1_SAdS}
\end{figure}

For the sake of simplicity, let us rewrite Eq.~(\ref{eq:function_B(r)_SAdS}) as
\begin{equation}
B(r)=B_{0}+\frac{B_{1}}{r}+B_{2}r+B_{3}r^{2}.
\label{eq:function_B(r)_SAdS_2}
\end{equation}
Thus, the event horizons are the solutions of the surface equation given by
\begin{equation}
B(r)=\frac{B_{3}}{r}(r^{3}+a_{2}r^{2}+a_{1}r+a_{0})=0,
\label{eq:surface_equation_SAdS}
\end{equation}
where
\begin{equation}
a_{2}=\frac{B_{2}}{B_{3}},
\label{eq:a2_SAdS}
\end{equation}
\begin{equation}
a_{1}=\frac{B_{0}}{B_{3}},
\label{eq:a1_SAdS}
\end{equation}
\begin{equation}
a_{0}=\frac{B_{1}}{B_{3}}.
\label{eq:a0_SAdS}
\end{equation}
Its solutions are given by \cite{NIST:2010}
\begin{equation}
r_{1}=-\frac{1}{3}a_{2}+\frac{1}{3}(I+J),
\label{eq:r1_SAdS}
\end{equation}
\begin{equation}
r_{2}=-\frac{1}{3}a_{2}+\frac{1}{3}(\rho I+\rho^{2} J),
\label{eq:r2_SAdS}
\end{equation}
\begin{equation}
r_{3}=-\frac{1}{3}a_{2}+\frac{1}{3}(\rho J+\rho^{2} I),
\label{eq:r3_SAdS}
\end{equation}
where
\begin{equation}
I=\sqrt[3]{-\frac{27}{2}v+\frac{3}{2}\sqrt{-3d}},
\label{eq:I_NIST}
\end{equation}
\begin{equation}
J=-\frac{3y}{I},
\label{eq:J_NIST}
\end{equation}
\begin{equation}
\rho=\mbox{e}^{2\pi i/3},
\label{eq:rho_NIST}
\end{equation}
\begin{equation}
\rho^{2}=\mbox{e}^{-2\pi i/3},
\label{eq:rho2_NIST}
\end{equation}
with
\begin{equation}
v=\frac{1}{27}(2a_{2}^{3}-9a_{2}a_{1}+27a_{0}),
\label{eq:v_NIST}
\end{equation}
\begin{equation}
d=-4y^{3}-27v^{2},
\label{eq:d_NIST}
\end{equation}
\begin{equation}
y=\frac{1}{3}(3a_{1}-a_{2}^{2}).
\label{eq:y_NIST}
\end{equation}
Therefore, we can rewrite the function $B(r)$ as
\begin{equation}
B(r)=\frac{B_{3}}{r}(r-r_{1})(r-r_{2})(r-r_{3}).
\label{eq:function_B(r)_SAdS_3}
\end{equation}
In this representation, the only positive real root is $r_{1}$, which corresponds to the exterior event horizon. The behavior of the exterior event horizon $r_{1}$ is shown in Fig.~\ref{fig:Fig1_SAdS}. In the limit when $\tilde{\Lambda} \rightarrow 0$, the complex roots $r_{2}$ and $r_{3}$ go to infinity and hence they decouple from the general solution.

For the chosen values, the exterior event horizon is at $r_{1}=1.74401$. The (unphysical) apparent event horizons are located at $r_{2}=-0.62201+5.31814i$ and $r_{3}=-0.62201-5.31814i$. It is worth noticing that the parameter $B_{3}$ must be non-zero, i.e., $\tilde{\Lambda} \neq 0$, which means that the term playing the role of an effective cosmological constant necessarily contributes to the energy density of such a spacetime.

In the next section, we will study the behavior of quantum scalar particles propagating outside the exterior event horizon of the SAdS black hole with an f(R) global monopole.

The outline of this paper is the following. In Section \ref{KG_equation}, we discuss the separability of the Klein-Gordon equation and then solve it in terms of the Heun functions. In Section \ref{Hawking_radiation}, we examine the Hawking radiation of scalar particles. In Section \ref{Quasibound_states}, we investigate the quasibound states by obtaining the spectrum of resonant frequencies. In Section \ref{Wave_functions}, we analyze the radial wave eigenfunctions. Finally, in Section \ref{Summary}, we present our concluding remarks. In \ref{Appendix}, we present the general Heun functions, as well as its deformed case.
%
%
\section{Klein-Gordon equation}\label{KG_equation}
In order to discuss the motion of quantum scalar particles propagating in a curved spacetime, we will consider the covariant Klein-Gordon equation with a non-minimal coupling, which is given by
\begin{equation}
\biggl\{\frac{1}{\sqrt{-g}}\partial_{\sigma}(g^{\sigma\tau}\sqrt{-g}\partial_{\tau})-(\mu^{2}+\xi R)\biggl\}\Psi(\mathbf{r})=0,
\label{eq:Klein-Gordon_Ricci}
\end{equation}
where $\mu$ is the mass of the scalar particle, and $\xi$ is the coupling constant. In the background under consideration, the Ricci curvature scalar $R$ is given by
\begin{eqnarray}
R & = & -\frac{1}{r^{2}}\biggl[r^{2}\frac{d^{2}B(r)}{dr^{2}}+4r\frac{dB(r)}{dr}+2B(r)-2\biggr]\nonumber\\
  & = & -12B_{3}-\frac{2(B_{0}-1)}{r^{2}}-\frac{6B_{2}}{r}.
\label{eq:Ricci_curvature}
\end{eqnarray}
Thus, by substituting the metric (\ref{eq:spherically_symmetric_f(R)}) into the Klein-Gordon equation (\ref{eq:Klein-Gordon_Ricci}), we get
\begin{eqnarray}
&& \biggl\{-\frac{r^{2}}{B(r)}\frac{\partial^{2}}{\partial t^{2}}+\frac{\partial}{\partial r}\biggl[r^{2}B(r)\frac{\partial}{\partial r}\biggr]-(\mu^{2}+\xi R)r^{2}\nonumber\\
&& +\frac{\partial^{2}}{\partial \theta^{2}}+\cot\theta\frac{\partial}{\partial \theta}+\csc^{2}\theta\frac{\partial^{2}}{\partial \phi^{2}}\biggr\}\Psi(t,r,\theta,\phi)=0.
\label{eq:mov_SAdS}
\end{eqnarray}

Now, we need to choose a suitable separation for the dependent variables of	 the scalar wave function $\Psi(t,r,\theta,\phi)$. Due to the spherical symmetry, we will write the scalar wave function as
\begin{equation}
\Psi(t,r,\theta,\phi)=\mbox{e}^{-i \omega t}u(r)Y_{lm}(\theta,\phi),
\label{eq:wave_function}
\end{equation}
where $\omega$ is the frequency (energy) of the scalar particle, $Y_{lm}(\theta,\phi)$ is the spherical harmonic function, and $u(r)=U(r)/r$ is the radial function. Thus, Eq.~(\ref{eq:mov_SAdS}) is separated in two parts, namely,
\begin{eqnarray}
&& \frac{1}{\sin^{2}\theta}\frac{\partial^{2} Y_{lm}(\theta,\phi)}{\partial \phi^{2}}\nonumber\\
&& +\frac{1}{\sin\theta}\frac{\partial}{\partial \theta}\biggl[\sin\theta\frac{\partial Y_{lm}(\theta,\phi)}{\partial \theta}\biggr]=0
\label{eq:angular_SAdS}
\end{eqnarray}
and
\begin{eqnarray}
&& \frac{d^{2} U(r)}{d r^{2}}+\frac{1}{B(r)}\frac{d B(r)}{d r}\frac{d U(r)}{d r}\nonumber\\
&& +\biggl\{\frac{\omega^{2}}{[B(r)]^{2}}-\frac{1}{r^{2}B(r)}\biggl[\lambda_{lm}+(\mu^{2}+\xi R)r^{2}\nonumber\\
&& +r\frac{d B(r)}{d r}\biggr]\biggr\}U(r)=0,
\label{eq:radial_SAdS}
\end{eqnarray}
where $\lambda_{lm}=l(l+1)$ is a separation constant, with $l$ being the azimuthal quantum number. In what follows, we will discuss and solve the radial part.
%
%
\subsection{Effective potential}
\begin{figure}
		\includegraphics[width=0.45\textwidth]{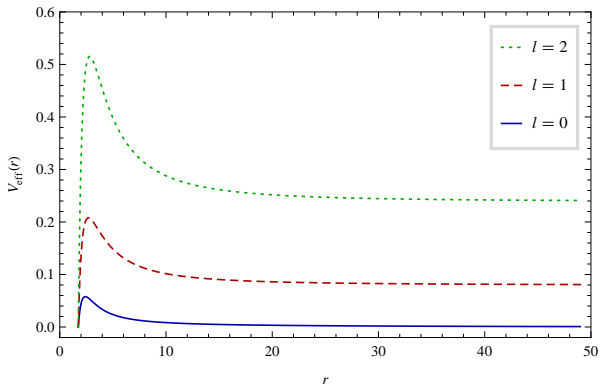}
	\caption{The effective potentials for $\mu=0$, $\xi=1/6$, and different values of the azimuthal quantum number $l(=0,1,2)$.}
	\label{fig:Fig2_SAdS}
\end{figure}

At this point, we would like to analyze the behavior of the effective potential, $V_{eff}(r)$. The radial equation given by Eq.~(\ref{eq:radial_SAdS}) can be written as
\begin{equation}
\frac{d^{2} U(r)}{d r_{*}^{2}}+[\omega^{2}-V_{eff}(r)]U(r)=0,
\label{eq:radial_SAdS_tortoise}
\end{equation}
where
\begin{equation}
V_{eff}(r)=B(r)\biggl[\frac{\lambda_{lm}}{r^{2}}+\mu^{2}+\xi R+\frac{1}{r}\frac{d B(r)}{d r}\biggr].
\label{eq:effective_potential_SAdS}
\end{equation}
As we can see, Eq.~(\ref{eq:radial_SAdS_tortoise}) looks like an one-dimensional Schr\"{o}dinger equation, where we have introduced the \textit{tortoise coordinate} $r_{*}$ defined by $dr_{*}=dr/B(r)$. The behavior of the effective potential $V_{eff}(r)$ is shown in Fig.~\ref{fig:Fig2_SAdS}, for some values of the azimuthal quantum number.

From Eq.~(\ref{eq:effective_potential_SAdS}) and Fig.~\ref{fig:Fig2_SAdS}, we see that the effective potential approaches to zero only in the case when $l=0$; it crosses the $r$-axis at $r=100.002$. For $l > 0$, the effective potential approaches to a (finite) constant as $r$ goes to infinity; it tends to $r=0.079$ for $l=1$, and to $r=0.239$ for $l=2$.
%
%
%
%
\subsection{Radial equation}
Now, let us solve the radial part of the Klein-Gordon equation. To do this, we substitute the function $B(r)$, given by Eq.~(\ref{eq:function_B(r)_SAdS_3}), into the radial equation, given by Eq.~(\ref{eq:radial_SAdS}), in order to get
\begin{eqnarray}
&& \frac{d^{2} U(r)}{d r^{2}}+\biggl(-\frac{1}{r}+\frac{1}{r-r_{1}}+\frac{1}{r-r_{2}}+\frac{1}{r-r_{3}}\biggr)\frac{d U(r)}{d r}\nonumber\\
&& +\biggl\{\frac{1}{3r^{2}B_{3}(r-r_{1})(r-r_{2})(r-r_{3})}\{r(B_{0}-3\lambda_{lm}-1)\nonumber\\
&& +3r^{2}[B_{2}+B_{3}(r_{1}+r_{2}+r_{3})]-3r^{3}\mu^{2}-3B_{3}r_{1}r_{2}r_{3}\}\nonumber\\
&& +\frac{r^{2}\omega^{2}}{B_{3}^{2}(r-r_{1})^{2}(r-r_{2})^{2}(r-r_{3})^{2}}\biggr\}U(r)=0,
\label{eq:radial_r_SAdS}
\end{eqnarray}
where we have chosen the conformal coupling ($\xi=1/6$), and used the Ricci curvature scalar given by Eq.~(\ref{eq:Ricci_curvature}).

Equation (\ref{eq:radial_r_SAdS}) seems to be a Fuchsian second-order equation with four finite regular singularities at the points $r=(0,r_{j})$, with $j=1,2,3$. Thus, it can be transformed into a kind of Heun equation. To do this, we have to define a new radial coordinate, $z$, by using the following homographic substitution
\begin{equation}
z=\frac{r-r_{1}}{r}\zeta,
\label{eq:coordinate_z_SAdS}
\end{equation}
where
\begin{equation}
\zeta=\frac{r_{2}}{r_{2}-r_{1}}.
\label{eq:zeta_SAdS}
\end{equation}
This transformation brings the singularities ($r_{1}$,$r_{2}$,$r_{3}$) to the points ($0$,$1$,$b$), where the singularity parameter $b$ is given by
\begin{equation}
b=\frac{r_{3}-r_{1}}{r_{3}}\zeta.
\label{eq:z1_SAdS}
\end{equation}
In addition, this transformation gives two important limits: when $r \rightarrow r_{1}$ implies that $z \rightarrow 0$, and when $r \rightarrow \infty$ implies that $z \rightarrow 1$. It means that we will obtain a solution which is analytical in the range $0 \leq z \leq 1$, that is, valid for $r_{1} \leq r \leq \infty$. Therefore, it totally agrees with the fact that we are interested on the motion of scalar particles propagating outside the exterior event horizon.

Thus, by substituting Eq.~(\ref{eq:coordinate_z_SAdS}) into Eq.~(\ref{eq:radial_r_SAdS}), we obtain
\begin{eqnarray}
&& \frac{d^{2} U(z)}{d z^{2}}+\biggl(\frac{1}{z}+\frac{1}{z-1}+\frac{1}{z-b}\biggr)\frac{d U(z)}{d z}\nonumber\\
&& +\biggl[\sum_{j=0}^{3}\frac{L_{j}}{z-z_{j}}+\sum_{j=0}^{2}\frac{Q_{j}}{(z-z_{j})^{2}}\nonumber\\
&& -\frac{\mu^{2}}{B_{3}(z-\zeta)^{2}}\biggr]U(z)=0,
\label{eq:radial_z_SAdS}
\end{eqnarray}
where $j=0,1,2,3$ labels the singularities $z=0,1,b,\zeta$. The parameter $\zeta$ is called an apparent singularity, since it can be removed (or have its power reduced) by performing some suitable transformations. Thus, Eq.~(\ref{eq:radial_z_SAdS}) is almost a Heun-type, where we just need to remove (or reduce the power) that apparent singularity. It is easy to see that there are two ways to do this: the simplest one is set $\mu=0$, and the other is to choose $\mu^{2} \propto B_{3}$ together with a specific transformation of the dependent variable $U(z)$. In fact, this was first noted by Kraniotis \cite{JPhysCommun.3.035026}, when he studied the massive Dirac equation in Kerr-Newman black hole spacetimes. Therefore, in what follows, we will solve the covariant Klein-Gordon equation with a conformal coupling in the SAdS black hole with an f(R) global monopole for both massless and massive scalar particles.
%
%
\subsection{Case 1: massless scalar particles}
For massless scalar particles $(\mu=0)$, we define a new dependent variable $U$ by performing the following \textit{F-homotopic} transformation
\begin{equation}
U(z)=z^{A_{1}}(z-1)^{A_{2}}(z-b)^{A_{3}}Z(z),
\label{eq:F-homotopic_radial_z_SAdS_Case1}
\end{equation}
where the exponents $A_{j}$ obey to the following indicial equation
\begin{equation}
F(s)=s(s-1)+s+Q_{j}=s^{2}+Q_{j}=0,
\label{eq:indicial_equation}
\end{equation}
whose roots are given by
\begin{equation}
s_{1,2}^{z=0}=\pm i\frac{(\zeta-1)(\zeta-b)\omega}{B_{3} r_{1} b} \equiv A_{1},
\label{eq:A1_radial_z_SAdS_Case1}
\end{equation}
\begin{equation}
s_{1,2}^{z=1}=\pm i\frac{(\zeta-1)(\zeta-b)\omega}{B_{3} r_{1} (b-1)} \equiv A_{2},
\label{eq:A2_radial_z_SAdS_Case1}
\end{equation}
\begin{equation}
s_{1,2}^{z=b}=\pm i\frac{(\zeta-1)(\zeta-b)\omega}{B_{3} r_{1} b (b-1)} \equiv A_{3}.
\label{eq:A3_radial_z_SAdS_Case1}
\end{equation}
Thus, by substituting Eqs.~(\ref{eq:F-homotopic_radial_z_SAdS_Case1})-(\ref{eq:A3_radial_z_SAdS_Case1}) into Eq.~(\ref{eq:radial_z_SAdS}), we get
\begin{eqnarray}
&& \frac{d^{2} Z(z)}{d z^{2}}+\biggl(\frac{1+2A_{1}}{z}+\frac{1+2A_{2}}{z-1}+\frac{1+2A_{3}}{z-b}\biggr)\frac{d Z(z)}{d z}\nonumber\\
&& +\frac{A_{4}z-A_{5}}{z(z-1)(z-b)}Z(z)=0,
\label{eq:radial_z_SAdS_Case1}
\end{eqnarray}
where the coefficients $A_{4}$, and $A_{5}$ are given by
\begin{eqnarray}
A_{4} & = & 2(A_{1}+A_{2}+A_{3}+A_{1}A_{2}+A_{1}A_{3}+A_{2}A_{3})\nonumber\\
&& -\frac{B_{2} (b+1-\zeta)}{B_{3} \zeta  r_{1}}\nonumber\\
&& -\frac{(b+1-2 \zeta) [(2 \zeta -1) b-2 (\zeta -1) \zeta ]}{(\zeta -1) \zeta  (b-\zeta )}\nonumber\\
&& -\frac{2 (\zeta -1)^2 \omega ^2 [(b-1) b+1] (b-\zeta )^2}{B_{3}^2 r_{1}^2 (b-1)^2 b^2},
\label{eq:A4_radial_z_SAdS_Case1}
\end{eqnarray}
\begin{eqnarray}
A_{5} & = & A_{1}+A_{3}+A_{1}A_{3}+(A_{1}+A_{2}+2A_{1}A_{2})b\nonumber\\
&& -2 (\zeta -1)+\biggl(\frac{1}{\zeta }-2\biggr) b\nonumber\\
&& +\frac{(\zeta -1) (b-\zeta ) (B_{0}+3 B_{2} r_{1}-3 \lambda_{lm} -1)}{3 B_{3} \zeta  r_{1}^2}\nonumber\\
&& -\frac{2 (\zeta -1)^2 \omega ^2 (b+1) (b-\zeta )^2}{B_{3}^2 r_{1}^2 b^2}.
\label{eq:A5_radial_z_SAdS_Case1}
\end{eqnarray}
The massless radial equation, given by Eq.~(\ref{eq:radial_z_SAdS_Case1}), is similar to the general Heun equation (see Eq.~(\ref{eq:canonical_form_general_Heun}) in \ref{Appendix}). Therefore, its analytical solution is given by
\begin{eqnarray}
U(z) & = & z^{A_{1}}(z-1)^{A_{2}}(z-b)^{A_{3}}\nonumber\\
&& \times \{C_{1}\ \mbox{HeunG}(b,q;\alpha,\beta,\gamma,\delta;z)\nonumber\\
&& + C_{2}\ z^{1-\gamma}\ \mbox{HeunG}(b,q_{2};\alpha_{2},\beta_{2},\gamma_{2},\delta;z)\},
\label{eq:general_solution_radial_z_SAdS_Case1}
\end{eqnarray}
where $C_{1}$ and $C_{2}$ are constants (to be determined). The parameters $\alpha$, $\beta$, $\gamma$, $\delta$, $\epsilon$, and $q$ are given by
\begin{equation}
\alpha=1+A_{1}+A_{2}+A_{3},
\label{eq:alpha_radial_z_SAdS_Case1}
\end{equation}
\begin{equation}
\beta=1+A_{1}+A_{2}+A_{3},
\label{eq:beta_radial_z_SAdS_Case1}
\end{equation}
\begin{equation}
\gamma=1+2A_{1},
\label{eq:gamma_radial_z_SAdS_Case1}
\end{equation}
\begin{equation}
\delta=1+2A_{2},
\label{eq:delta_radial_z_SAdS_Case1}
\end{equation}
\begin{equation}
\epsilon=1+2A_{3},
\label{eq:eta_radial_z_SAdS_Case1}
\end{equation}
\begin{equation}
q=A_{5}.
\label{eq:q_radial_z_SAdS_Case1}
\end{equation}
Furthermore, the auxiliary parameters $\alpha_{2}$, $\beta_{2}$, $\gamma_{2}$, and $q_{2}$ are given by
\begin{equation}
\alpha_{2}=\alpha+1-\gamma,
\label{eq:alpha_1_general_Heun_Case1}
\end{equation}
\begin{equation}
\beta_{2}=\beta+1-\gamma,
\label{eq:beta_1_general_Heun_Case1}
\end{equation}
\begin{equation}
\gamma_{2}=2-\gamma.
\label{eq:gamma_1_general_Heun_Case1}
\end{equation}
\begin{equation}
q_{2}=q+(\alpha\delta+\epsilon)(1-\gamma).
\label{eq:q_1_general_Heun_Case1}
\end{equation}
These are two linearly independent solutions of the general Heun equation since $\gamma$ is not a positive integer, and they correspond to the exponents $0$ and $1-\gamma$ at $z=0$. It is worth emphasizing that the final expressions for these parameters depend on the signs to be chosen for the exponents $A_{j}$, which are given by Eqs.~(\ref{eq:A1_radial_z_SAdS_Case1})-(\ref{eq:A3_radial_z_SAdS_Case1}).
%
%
\subsection{Case 2: massive scalar particles}
For massive scalar particles $(\mu=\sqrt{3B_{3}}/2)$, the dependent variable $U$ is now transformed as
\begin{equation}
U(z)=z^{A_{1}}(z-1)^{A_{2}}(z-b)^{A_{3}}\frac{Z(z)}{(z-\zeta)^{\frac{1}{2}}},
\label{eq:F-homotopic_radial_z_SAdS_Case2}
\end{equation}
where the coefficients $A_{1}$, $A_{2}$, and $A_{3}$ are the same as for the Case 1, that is, theu are given by Eqs.~(\ref{eq:A1_radial_z_SAdS_Case1})-(\ref{eq:A3_radial_z_SAdS_Case1}). Thus, by substituting Eq.~(\ref{eq:F-homotopic_radial_z_SAdS_Case2}) into Eq.~(\ref{eq:radial_z_SAdS}), we get
\begin{eqnarray}
&& \frac{d^{2} Z(z)}{d z^{2}}+\biggl(\frac{1+2A_{1}}{z}+\frac{1+2A_{2}}{z-1}+\frac{1+2A_{3}}{z-b}\nonumber\\
&& -\frac{1}{z-\zeta}\biggr)\frac{d Z(z)}{d z}+\biggl[\frac{-D_{1}-D_{2}-D_{3}+D_{3}\zeta+D_{2}b}{(z-1)(z-b)}\nonumber\\
&& +\frac{(D_{1}+D_{3}-D_{3}\zeta)b}{z(z-1)(z-b)}\nonumber\\
&& +\frac{D_{3}(\zeta-1)\zeta}{z(z-1)(z-\zeta)}\biggr]Z(z)=0,
\label{eq:radial_z_SAdS_Case2}
\end{eqnarray}
where the coefficients $D_{1}$, $D_{2}$, and $D_{3}$ are given by
\begin{eqnarray}
D_{1} & = & -\frac{2 A_{1} A_{3}+A_{1}+A_{3}}{b}\nonumber\\
&& -\frac{4 A_{1} A_{2} \zeta +2 A_{1} \zeta -2 A_{1}+2 A_{2} \zeta -2 \zeta  L_{1}-1}{2 \zeta },
\label{eq:D1_radial_z_SAdS_Case2}
\end{eqnarray}
\begin{eqnarray}
D_{2} & = & L_{3}+\frac{2 A_{2} A_{3}+A_{2}+A_{3}}{b-1}+\frac{2 A_{1} A_{3}+A_{1}+A_{3}}{b}\nonumber\\
&& -\frac{2 A_{3}+1}{2 (b-\zeta )},
\label{eq:D2_radial_z_SAdS_Case2}
\end{eqnarray}
\begin{eqnarray}
D_{3} & = & \frac{2 A_{3}+1}{2 (b-\zeta )}+\frac{-2 A_{1} \zeta +2 A_{1}-2 A_{2} \zeta -2 \zeta +2 \zeta ^2 L_{4}}{2 (\zeta -1) \zeta }\nonumber\\
&& +\frac{1-2 \zeta  L_{4}}{2 (\zeta -1) \zeta },
\label{eq:D3_radial_z_SAdS_Case2}
\end{eqnarray}
with
\begin{eqnarray}
L_{1} & = & \frac{5 (\zeta -1) \zeta -5 \zeta  b+b}{4 \zeta  b}\nonumber\\
&& -\frac{(\zeta -1) (b-\zeta ) (B_{0}+3 B_{2} r_{1}-3 \lambda_{lm} -1)}{3 B_{3} \zeta  r_{1}^2 b}\nonumber\\
&& +\frac{2 (\zeta -1)^2 \omega ^2 (b+1) (b-\zeta )^2}{B_{3}^2 r_{1}^2 b^3},
\label{eq:L1_radial_z_SAdS_Case2}
\end{eqnarray}
\begin{eqnarray}
L_{2} & = & \frac{(5 \zeta-4) b+4-5 \zeta ^2}{(4 \zeta-4) b+4-4 \zeta}\nonumber\\
&& +\frac{(b-\zeta ) [B_{0} (\zeta -1)+3 \zeta B_{2} r_{1}]}{3 B_{3} \zeta  r_{1}^2 (b-1)}\nonumber\\
&& +\frac{(1-\zeta)(3 \lambda_{lm} +1)]}{3 B_{3} \zeta  r_{1}^2 (b-1)}\nonumber\\
&& -\frac{2 (\zeta -1)^2 \omega ^2 (b-2) (b-\zeta )^2}{B_{3}^2 r_{1}^2 (b-1)^3},
\label{eq:L2_radial_z_SAdS_Case2}
\end{eqnarray}
\begin{eqnarray}
L_{3} & = & \frac{4 (b-1) b-5 (\zeta -1) \zeta }{4 (b-1) b (b-\zeta )}\nonumber\\
&& +\frac{(\zeta -1) \zeta  (B_{0}+3 B_{2} r_{1}-3 \lambda_{lm} -1)}{3 B_{3} \zeta  r_{1}^2 (b-1) b}\nonumber\\
&& +\frac{(\zeta -1) b (3 \lambda_{lm} +1-B_{0})}{3 B_{3} \zeta  r_{1}^2 (b-1) b}\nonumber\\
&& -\frac{2 (\zeta -1)^2 \omega ^2 (2 b-1) (b-\zeta )^2}{B_{3}^2 r_{1}^2 (b-1)^3 b^3},
\label{eq:L3_radial_z_SAdS_Case2}
\end{eqnarray}
\begin{eqnarray}
L_{4} & = & \frac{1}{4} \biggl(\frac{1}{1-\zeta }+\frac{1}{b-\zeta }-\frac{4 B_{2}}{B_{3} r_{1} \zeta}-\frac{1}{\zeta}\biggr).
\label{eq:L4_radial_z_SAdS_Case2}
\end{eqnarray}
The massive radial equation, given by Eq.~(\ref{eq:radial_z_SAdS_Case2}), is similar to the deformed Heun equation (see Eq.~(\ref{eq:canonical_deformed_general_Heun}) in \ref{Appendix}), where $\zeta$ plays the role of an apparent singularity. Therefore, its analytical solution is given by
\begin{eqnarray}
U(z) & = & z^{A_{1}}(z-1)^{A_{2}}(z-b)^{A_{3}}(z-\zeta)^{-\frac{1}{2}}\nonumber\\
&& \times \{C_{1}\ \mbox{HeunG}(b,q;\alpha,\beta,\gamma,\delta;z)\nonumber\\
&& + C_{2}\ z^{1-\gamma}\ \mbox{HeunG}(b,q_{2};\alpha_{2},\beta_{2},\gamma_{2},\delta;z)\},
\label{eq:general_solution_radial_z_SAdS_Case2}
\end{eqnarray}
where $C_{1}$ and $C_{2}$ are constants (to be determined). In this case, the parameters $\alpha$, $\beta$, $\gamma$, $\delta$, $\epsilon$, and $q$ are now given by
\begin{eqnarray}
\alpha & = & 1+A_{1}+A_{2}+A_{3}\nonumber\\
&& -\frac{1}{2}\biggl[2(2 A_{1}^2+2 A_{1}+2 A_{2}^2+2 A_{2}+2 A_{3}^2+2 A_{3}\nonumber\\
&& +2 L_{1}-2 L_{3} b+2 L_{3}-2 \zeta  L_{4}+2 L_{4}+5)\biggr]^{\frac{1}{2}},
\label{eq:alpha_radial_z_SAdS_Case2}
\end{eqnarray}
\begin{eqnarray}
\beta & = & 1+A_{1}+A_{2}+A_{3}\nonumber\\
&& -\frac{1}{2}\biggl[2(2 A_{1}^2+2 A_{1}+2 A_{2}^2+2 A_{2}+2 A_{3}^2+2 A_{3}\nonumber\\
&& +2 L_{1}-2 L_{3} b+2 L_{3}-2 \zeta  L_{4}+2 L_{4}+5)\biggr]^{\frac{1}{2}},
\label{eq:beta_radial_z_SAdS_Case2}
\end{eqnarray}
\begin{equation}
\gamma=2+2A_{1},
\label{eq:gamma_radial_z_SAdS_Case2}
\end{equation}
\begin{equation}
\delta=2+2A_{2},
\label{eq:delta_radial_z_SAdS_Case2}
\end{equation}
\begin{equation}
\epsilon=1+2A_{3},
\label{eq:eta_radial_z_SAdS_Case2}
\end{equation}
\begin{equation}
q=-(D_{1}+D_{3}-D_{3}\zeta)b,
\label{eq:q_radial_z_SAdS_Case2}
\end{equation}
It is worth emphasizing that we have already added the unitary shifting to the parameters $\gamma$ and $\delta$, as described in \ref{Appendix}, as well as that the final expressions for these parameters also depend on the signs to be chosen for the exponents $A_{j}$, which are given by Eqs.~(\ref{eq:A1_radial_z_SAdS_Case1})-(\ref{eq:A3_radial_z_SAdS_Case1}). The auxiliary parameters $\alpha_{2}$, $\beta_{2}$, $\gamma_{2}$, and $q_{2}$ are given by the same relations as for the Case 1, that is, they are given by Eqs.~(\ref{eq:alpha_1_general_Heun_Case1})-(\ref{eq:q_1_general_Heun_Case1}).

Next, we will use these analytical solutions of the radial equation, in the SAdS black hole with an f(R) global monopole, and some properties of the general Heun functions to discuss some interesting physical phenomena, namely, the Hawking radiation, the spectrum of quasibound state frequencies and its corresponding wave eigenfunctions.
%
%
\section{Hawking radiation}\label{Hawking_radiation}
In order to discuss the Hawking radiation, we will obtain the wave solutions describing quantum scalar particles near the exterior event horizon of a SAdS black hole with an f(R) global monopole. To do this, first we need to choose the signs of the exponents $A_{j}$ given by Eqs.~(\ref{eq:A1_radial_z_SAdS_Case1})-(\ref{eq:A3_radial_z_SAdS_Case1}); the negative sign is the correct choice, which will be proved in the discussion of the quasibound states.

In the limit when $r \rightarrow r_{1}$, which implies that $z \rightarrow 0$, we can evaluate the corresponding Heun functions from the expansion given by Eq.~(\ref{eq:serie_HeunG_todo_z}), and hence we get $\mbox{HeunG}(b,q;\alpha,\beta,\gamma,\delta;0) \sim 1$. Thus, the radial solutions for the Cases 1 and 2, which are given by Eqs.~(\ref{eq:general_solution_radial_z_SAdS_Case1}) and (\ref{eq:general_solution_radial_z_SAdS_Case2}), respectively, have the (same) asymptotic behavior at the exterior event horizon given by
\begin{equation}
u(r) \sim C_{1}\ (r-r_{1})^{A_{1}}+C_{2}\ (r-r_{1})^{-A_{1}},
\label{eq:exp_0_solucao_general_solution_radial_z_SAdS}
\end{equation}
where all remaining constants were included in $C_{1}$ and $C_{2}$. In fact, this algebraic expression is the same for the Cases 1 and 2, but the constants $C_{1}$ and $C_{2}$ have different contents in each case. Note that we recovered the original radial coordinate $r$, as well as the original radial function $u(r)$.

Now, by taking into account the contribution of the time coordinate, on the exterior surface of the SAdS black hole with an f(R) global monopole, the full wave solution can be written as
\begin{equation}
\Psi(r,t) \sim C_{1}\ \Psi_{\rm in}+C_{2}\ \Psi_{\rm out},
\label{eq:sol_onda_radial_z_SAdS}
\end{equation}
where the solutions describing the ingoing and outgoing scalar waves are given, respectively, by
\begin{equation}
\Psi_{\rm in}(r>r_{1})=\mbox{e}^{-i \omega t}(r-r_{1})^{-\frac{i}{2\kappa_{1}}\omega}
\label{eq:sol_out_2_SAdS}
\end{equation}
and
\begin{equation}
\Psi_{\rm out}(r>r_{1})=\mbox{e}^{-i \omega t}(r-r_{1})^{\frac{i}{2\kappa_{1}}\omega}.
\label{eq:sol_in_1_SAdS}
\end{equation}
The gravitational acceleration on the exterior horizon, $\kappa_{1}$, is defined as
\begin{equation}
\kappa_{1} \equiv \frac{1}{2} \left.\frac{dB(r)}{dr}\right|_{r=r_{1}} = \frac{B_{3}(r_{1}-r_{2})(r_{1}-r_{3})}{2r_{1}},
\label{eq:acel_grav_ext_SAdS}
\end{equation}
such that, from Eq.~(\ref{eq:A1_radial_z_SAdS_Case1}), we get
\begin{equation}
A_{1}=-\frac{i}{2\kappa_{1}}\omega.
\label{eq:beta/2_solucao_geral_radial_z_SAdS}
\end{equation}

Therefore, we follow the method described by Vieira \textit{et al.} \cite{AnnPhys.350.14} to compute the relative scattering probability, $\Gamma_{1}$, and the Hawking radiation spectra, $\bar{N}_{\omega}$. They are given by
\begin{equation}
\Gamma_{1}=\left|\frac{\Psi_{out}(r>r_{1})}{\Psi_{out}(r<r_{1})}\right|^{2}=\mbox{e}^{-\frac{2\pi}{\kappa_{1}}\omega},
\label{eq:taxa_refl_SAdS}
\end{equation}
and
\begin{equation}
\bar{N}_{\omega}=\frac{\Gamma_{1}}{1-\Gamma_{1}}=(\mbox{e}^{\frac{2\pi}{\kappa_{1}}\omega}-1)^{-1}.
\label{eq:espectro_rad_SAdS}
\end{equation}
From these results, we conclude that the Hawking radiation spectrum, for both massless and massive scalar particles in the SAdS black hole with an f(R) global monopole, is analogous to the black body spectrum, which has a thermal character. It is worth noticing that we used the definition of the Hawking temperature given by $k_{B}T_{+}=\hbar\kappa_{+}/2\pi$, where $k_{B}$ is the well know Boltzmann constant.

These results were obtained from the analytical solutions of the Klein-Gordon equation in the background under consideration. In fact, that is a semi-classical field theory approach.
%
%
\section{Quasibound states}\label{Quasibound_states}
The quasibound states, also known as quasistationary levels or resonance spectra, are solutions of the equation of motion that tend to zero far from the black hole at spatial infinity. This means that they are localized in the potential well of the black hole. Thus, that is a boundary value problem with two associated boundary conditions, which gives rise to a characteristic resonance equation for the frequency (energy) of the quantum particle.

In this physical phenomenon, the flux of quantum particles crosses into the black hole event horizon, by resulting in a spectrum that has complex frequencies, so that it is called a quasispectrum of resonant frequencies and expressed as $\omega=\omega_{R}+i\omega_{I}$, where $\omega_{R}$ and $\omega_{I}$ are the real and imaginary parts of the frequencies, respectively. The real part describes the oscillation frequency, while the imaginary part is related to the decay (if $\mbox{Im}[\omega] < 0$) or growth (if $\mbox{Im}[\omega] > 0$) rate with the time.

There are some different approaches used to derive the characteristic resonance equation \cite{PhysRevD.76.084001,PhysRevD.85.044031,PhysLettB.749.167,PhysRevD.103.044062}. In the present work, we will use the Vieira-Bezerra-Kokkotas method \cite{AnnPhys.373.28,PhysRevD.104.024035} to obtain the spectrum of quasibound state frequencies.

Thus, the first boundary condition is such that the radial solution should describe an ingoing wave at the exterior event horizon. Then, we have to impose that $C_{2}=0$ in Eq.~(\ref{eq:sol_onda_radial_z_SAdS}), as well as in Eqs.~(\ref{eq:general_solution_radial_z_SAdS_Case1}) and (\ref{eq:general_solution_radial_z_SAdS_Case2}). On the other hand, the second boundary condition is such that the radial solution should tend to zero far from the black hole at asymptotic infinity. In order to fully satisfy this condition, we have to take the limit $r \rightarrow \infty$ on the radial solutions given by Eqs.~(\ref{eq:general_solution_radial_z_SAdS_Case1}) and (\ref{eq:general_solution_radial_z_SAdS_Case2}), for the Cases 1 and 2, respectively. To do this, we will write these solutions in terms of the $\alpha$ and $\beta$ exponent solutions given by Eqs.~(\ref{eq:alpha_Heun_infinity}) and (\ref{eq:beta_Heun_infinity}). After some algebra, we get the following asymptotic behavior
\begin{equation}
u(r) \sim C_{1}\ \frac{1}{r}+C_{2}\ \frac{1}{r} \qquad\qquad \mbox{(Case 1)},
\label{eq:radial_infinity_SAdS_Case1}
\end{equation}
and
\begin{equation}
u(r) \sim C_{1}\ \frac{1}{\sqrt{r}}+C_{2}\ \frac{1}{\sqrt{r}} \qquad\ \, \mbox{(Case 2)}.
\label{eq:radial_infinity_SAdS_Case2}
\end{equation}
However, since $C_{2}=0$ from the first boundary condition, we have that
\begin{equation}
u(r) \sim C_{1}\ \frac{1}{r} \qquad\qquad\qquad\quad \mbox{(Case 1)},
\label{eq:radial_infinity_2_SAdS_Case1}
\end{equation}
and
\begin{equation}
u(r) \sim C_{1}\ \frac{1}{\sqrt{r}} \qquad\qquad\qquad\ \mbox{(Case 2)}.
\label{eq:radial_infinity_2_SAdS_Case2}
\end{equation}
Thus, the radial solutions given in terms of the general and deformed Heun functions tend to zero far from the black hole at asymptotic infinity, as required by the quasibound states.

Now, the final step is to use a matching procedure in order to bring the two different asymptotic regions into their common overlap region. To do this, we will use the polynomial condition of the Heun functions as described in the Vieira-Bezerra-Kokkotas method \cite{AnnPhys.373.28,PhysRevD.104.024035}, that is, we will obtain the spectrum of resonant frequencies by using the fact that the general Heun functions become a polynomial of degree $n$ if they satisfy the so-called $\alpha$-condition given by Eq.~(\ref{eq:alpha-condition}).
%
%
\subsection{Case 1: massless scalar particles}
\begin{figure}
		\includegraphics[width=0.45\textwidth]{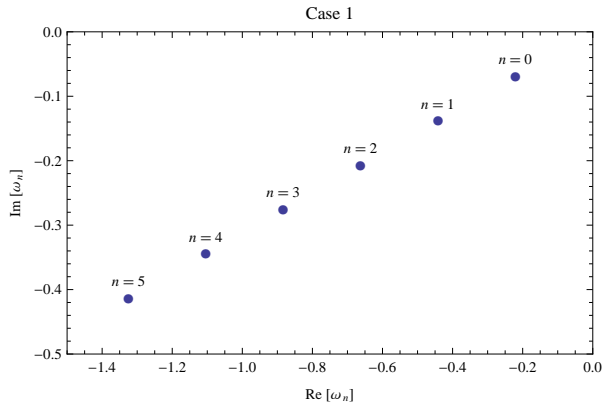}
	\caption{The massless scalar resonant frequencies $\omega_{n}$.}
	\label{fig:Fig3_SAdS_Case1}
\end{figure}

In this case, the parameter $\alpha$ is given by Eq.~(\ref{eq:alpha_radial_z_SAdS_Case1}), which can be simply written as
\begin{equation}
\alpha=1-2iE_{1}\omega,
\label{eq:alpha_omega_SAdS_Case1}
\end{equation}
where the coefficient $E_{1}$ is given by
\begin{equation}
E_{1}=\frac{(b-\zeta)(\zeta-1)}{B_{3}r_{1}(b-1)}=\frac{r_{2}}{B_{3}(r_{1}-r_{2})(r_{2}-r_{3})}.
\label{eq:E1_bound_states_SAdS}
\end{equation}
Note that the coefficient $E_{1}$ is a complex number ($E_{1} \in \mathbb{C}$). Then, by imposing the polynomial condition given by Eq.~(\ref{eq:alpha-condition}), we obtain the following expression for the massless scalar resonant frequencies
\begin{equation}
\omega_{n}=-i\frac{n+1}{2E_{1}},
\label{eq:omega_SAdS_Case1}
\end{equation}
where $n=0,1,2,\ldots$ is now the principal quantum number. Therefore, this is the spectrum of quasibound states for massless scalar particles propagating in the SAdS black hole with an f(R) global monopole. We shown some values of $\omega_{n}$ in Table \ref{tab:omega_SAdS_Case1}, and its behavior in Fig.~\ref{fig:Fig3_SAdS_Case1} as function of the principal quantum number $n$.

\begin{table}
\centering
\caption{Values of the massless scalar resonant frequencies $\omega_{n}$.}
\label{tab:omega_SAdS_Case1}
\begin{tabular*}{\columnwidth}{c@{\extracolsep{\fill}}ccccc@{}}
Case 1                        \\
\hline
$n$ & $\omega_{n}$            \\
\hline
0   & $-0.22077-0.06881i$ \\
1   & $-0.44155-0.13763i$ \\
10  & $-2.42853-0.75700i$ \\
\hline
\end{tabular*}
\end{table}

From Table \ref{tab:omega_SAdS_Case1} and Fig.~\ref{fig:Fig3_SAdS_Case1}, we see that the modulus of both real and imaginary parts of the massless scalar resonant frequencies increase with $n$, for fixed values of the parameters related to the f(R) global monopole. Therefore, the quasistationary levels consist of an infinite sequence of discrete levels, which are equally spaced. The imaginary part is always negative, which means damped oscillations and that the system may be stable.
%
%
\subsection{Case 2: massive scalar particles}
\begin{figure}
		\includegraphics[width=0.45\textwidth]{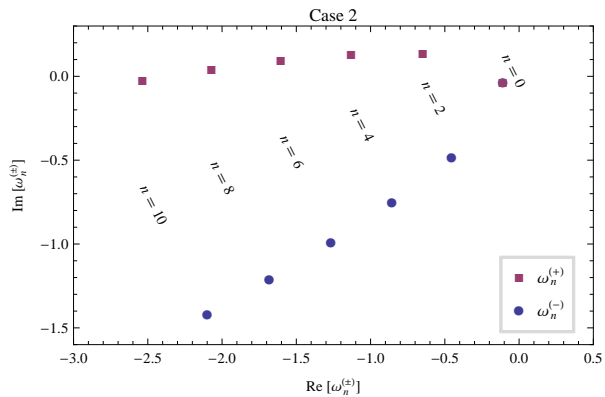}
	\caption{The massive scalar resonant frequencies $\omega_{n}^{(\pm)}$ for $\mu=0.17320$.}
	\label{fig:Fig4_SAdS_Case2}
\end{figure}

Now, let us analyze the case of massive scalar particles. In this case, the parameter $\alpha$ is given by Eq.~(\ref{eq:alpha_radial_z_SAdS_Case2}), which can be simply written as
\begin{equation}
\alpha=1-2iE_{1}\omega-\frac{1}{2}\sqrt{3-8iE_{1}\omega}.
\label{eq:alpha_omega_SAdS_Case2}
\end{equation}
Then, by imposing the polynomial condition given by Eq.~(\ref{eq:alpha-condition}), we obtain the following expressions for the massive scalar resonant frequencies
\begin{equation}
\omega_{n}^{(-)}=\frac{i(2n+1)-2\sqrt{n}}{4E_{1}}
\label{eq:omega1_SAdS_Case2}
\end{equation}
and
\begin{equation}
\omega_{n}^{(+)}=\frac{i(2n+1)+2\sqrt{n}}{4E_{1}}.
\label{eq:omega2_SAdS_Case2}
\end{equation}
This quasistationary levels are also complex, where $(\pm)$ labels the solutions; $(-)$ is the ``minus'' solution, while $(+)$ is the ``plus'' solution. Indeed, we obtained two solutions due to the fact that the $\alpha$-condition, in this case, leads to a second-order equation for $\omega$. Therefore, this is the spectrum of quasibound states for massive scalar particles propagating in the SAdS black hole with an f(R) global monopole. We shown some values of $\omega_{n}^{(\pm)}$ in Table \ref{tab:omega_SAdS_Case2}, and its behavior in Fig.~\ref{fig:Fig4_SAdS_Case2} as function of the principal quantum number $n$.

\begin{table}
\centering
\caption{Values of the massive scalar resonant frequencies $\omega_{n}^{(\pm)}$ for $\mu=0.17320$.}
\label{tab:omega_SAdS_Case2}
\begin{tabular*}{\columnwidth}{c@{\extracolsep{\fill}}ccccc@{}}
Case 2                                                \\
\hline
$n$ & $\omega_{n}^{(-)}$   & $\omega_{n}^{(+)}$   \\
\hline
0   & $-0.11038-0.03440i$ & $-0.11038-0.03440i$ \\
1   & $-0.26234-0.32400i$ & $-0.39998+0.11754i$ \\
10  & $-2.10050-1.42075i$ & $-2.53575-0.02445i$ \\
\hline
\end{tabular*}
\end{table}

From Table \ref{tab:omega_SAdS_Case2} and Fig.~\ref{fig:Fig4_SAdS_Case2}, we see that the modulus of both real and imaginary parts of the massive scalar resonant frequencies $\omega_{n}^{(-)}$ increase with $n$, for fixed values of the parameters related to the f(R) global monopole, and therefore we can conclude that $\omega_{n}^{(-)}$ are damped oscillations, and that the system may be stable. On the other hand, if the particles have resonant frequencies $\omega_{n}^{(+)}$, the system may present instability for some excited states ($1 \leq n \leq 9$), since the imaginary part of $\omega_{n}^{(+)}$ change its sign.

It is worth commenting that both massless and massive scalar resonant frequencies were obtained directly from the general Heun functions, by using a polynomial condition, and, to our knowledge, there is no similar result in the literature for the SAdS black hole with an f(R) global monopole.
%
%
\section{Wave eigenfunctions}\label{Wave_functions}
In order to analyze the wave eigenfunctions related to the massless and massive scalar resonant frequencies obtained in the previous section, we will use some properties of the general Heun functions and then obtain their polynomial expressions, which are presented in \ref{Appendix}.
%
%
\subsection{Case 1: massless scalar particles}
For massless scalar particles, the radial function $U(z)$ is given by Eq.~(\ref{eq:F-homotopic_radial_z_SAdS_Case1}). Thus, the radial wave eigenfunctions, for massless scalar particles propagating in the SAdS black hole with an f(R) global monopole, are given by
\begin{equation}
U_{n;s}(z)=C_{n;s}\ z^{A_{1}}(z-1)^{A_{2}}(z-b)^{A_{3}}\ \mbox{Hp}_{n;s}(z),
\label{eq:eigenfunctions_SAdS_Case1}
\end{equation}
where $C_{n;s}$ is a constant (to be determined). It is worth noticing that these radial wave eigenfunctions are degenerate, since the accessory parameter $q_{n;s}$ must be properly determined for each value of $s$, where $0 \leq s \leq n$.

\begin{figure}
		\includegraphics[width=0.45\textwidth]{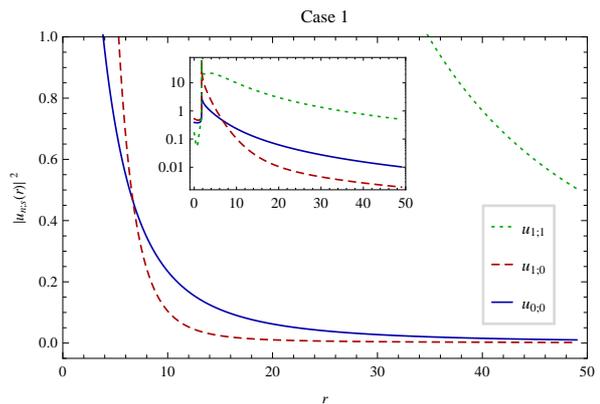}
	\caption{The first three squared massless radial wave eigenfunctions $u_{n;s}(r)=U_{n;s}(r)/r$ related to $\omega_{n}$. The units are in multiples of $C_{n;s}$.}
	\label{fig:Fig5_SAdS_Case1}
\end{figure}

Therefore, by using Eqs.~(\ref{eq:Hp_0,0}), (\ref{eq:Hp_1,0}), and (\ref{eq:Hp_1,1}), we can plot the first three squared massless radial wave eigenfunctions, which are presented in Fig.~\ref{fig:Fig5_SAdS_Case1}.

From Fig.~\ref{fig:Fig5_SAdS_Case1}, we see that the massless radial wave eigenfunctions present the desired behavior, that is, the decaying quasibound states (with $\mbox{Im}[\omega_{n}] < 0$) have a radial solution tending to zero at infinity and diverging at the exterior event horizon, so that it mathematically reaches a maximum value (see this in the log plot) and then crosses into the black hole.
%
%
\subsection{Case 2: massive scalar particles}
\begin{figure}
		\includegraphics[width=0.45\textwidth]{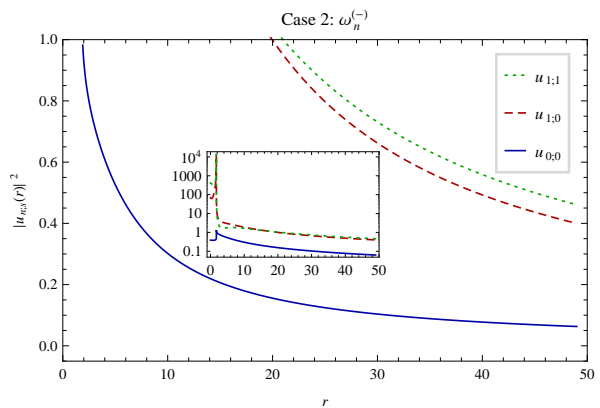}
	\caption{The first three squared massive radial wave eigenfunctions $u_{n;s}(r)=U_{n;s}(r)/r$ related to $\omega_{n}^{(-)}$ for $\mu=0.17320$. The units are in multiples of $C_{n;s}$.}
	\label{fig:Fig6_SAdS_Case2}
\end{figure}

\begin{figure}
		\includegraphics[width=0.45\textwidth]{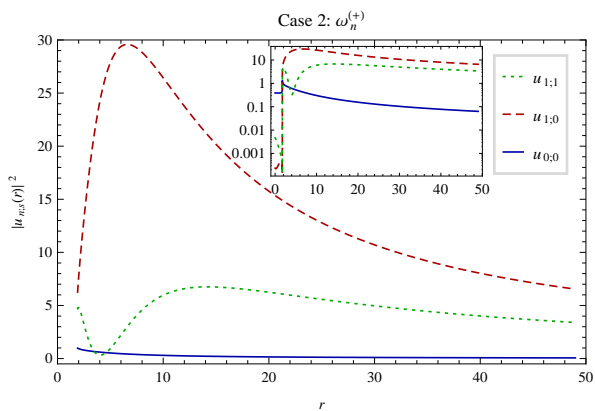}
	\caption{The first three squared massive radial wave eigenfunctions $u_{n;s}(r)=U_{n;s}(r)/r$ related to $\omega_{n}^{(+)}$ for $\mu=0.17320$. The units are in multiples of $C_{n;s}$.}
	\label{fig:Fig7_SAdS_Case2}
\end{figure}

For massive scalar particles propagating in the SAdS black hole with an f(R) global monopole, the radial function $U(z)$ is given by Eq.~(\ref{eq:F-homotopic_radial_z_SAdS_Case2}), so that we can write their radial wave eigenfunctions as
\begin{equation}
U_{n;s}(z)=C_{n;s}\ z^{A_{1}}(z-1)^{A_{2}}(z-b)^{A_{3}}\frac{\mbox{Hp}_{n;s}(z)}{(z-\zeta)^{\frac{1}{2}}},
\label{eq:eigenfunctions_SAdS_Case2}
\end{equation}
where $C_{n;s}$ is a constant (to be determined). Thus, by using Eqs.~(\ref{eq:Hp_0,0}), (\ref{eq:Hp_1,0}), and (\ref{eq:Hp_1,1}), we can show the first three squared massive radial wave eigenfunctions in Figs.~\ref{fig:Fig6_SAdS_Case2} and \ref{fig:Fig7_SAdS_Case2}.

From Fig.~\ref{fig:Fig6_SAdS_Case2}, we conclude that the massive scalar resonant frequencies $\omega_{n}^{(-)}$ describes quasibound states. On the other hand, in Fig.~\ref{fig:Fig7_SAdS_Case2} we can see that only massive scalar particles in the fundamental mode, with resonant frequencies $\omega_{0}^{(+)}$, are quasibound states. Otherwise, for $\omega_{n \geq 1}^{(+)}$, the radial solutions go to zero at the exterior event horizon and then they are not quasibound states.
%
%
\section{Final remarks}\label{Summary}
In this work, we presented analytical solutions for both angular and radial parts of the covariant Klein-Gordon equation with a conformal coupling in the SAdS black hole with an f(R) global monopole. The angular solution is given in terms of the spherical harmonic function. On the other hand, the radial solution is given in terms of the general and deformed Heun functions for massless and massive scalar fields, respectively.

We studied three very important physical phenomena related to the radial solution, namely, the Hawking radiation spectrum, which we found that is similar to the black body radiation, the resonant frequencies, where we imposed the boundary conditions related to the quasibound states, and the wave eigenfunctions, which describes the behavior of quantum scalar particles near the exterior event horizon and far from the black hole at the asymptotic infinity.

The resonant frequencies $\omega_{n}$ and $\omega_{n}^{(-)}$, which corresponds to the massless and massive scalar particles, respectively, have similar behavior, that is, their imaginary parts are always negative, do not change their signs, and hence they describes quasibound states in stable systems. On the other hand, the massive scalar resonant frequencies $\omega_{n}^{(+)}$ have a positive imaginary part in the fundamental mode ($n=0$), but their first nine excited modes ($1 \leq n \leq 9$) cross down the horizontal axis and then become negative, which may indicate some kind of phase transition and/or instability in the system.

It is worth calling attention to the fact that this quasistationary levels are associated with the interaction of quantum scalar fields and the curvature of the spacetime under consideration and therefore it is an very interesting semi-classical phenomena, which can give us some insights in the physics of black holes, and larger astrophysical systems as well, and for this reason should be investigated from a theoretical point of view. We hope that, in a near future, it may be used to fit some astrophysical data.

More generally, there has been considerable activity in recent years in the area of quantum gravity phenomenology, which seeks to find observational signatures of the quantum nature of spacetime. These studies may shed some light on the physics of black holes, and they can indicate a possible path to the construction of a quantum theory of gravity.

Finally, it is worth commenting that, in principle, we can use this approach to study quantum scalar fields propagating in a Schwarzschild-de Sitter (SdS) black hole spacetime. However, in such a case, there exist three event horizons, which means that we have to take into account the spatial region between the exterior event horizon and the cosmological horizon; it would be interesting to extend our analysis to this context. In fact, some preliminary investigations which concern this extension were already done, and we expect to publish some results in the near future.
%
%
\begin{acknowledgements}
H.S.V. is funded by the Alexander von Humboldt-Stiftung/Foundation (Grant No. 1209836). This study was financed in part by the Coordena\c c\~{a}o de Aperfei\c coamento de Pessoal de N\'{i}vel Superior - Brasil (CAPES) - Finance Code 001.
\end{acknowledgements}
%
%
\appendix
\section{The general and deformed Heun equations}\label{Appendix}
Here we present some features about the general Heun equation and its deformed case.
%
%
\subsection{The general Heun equation}
The Heun equation, also called general Heun equation, is a Fuchsian type, with regular singularities at $z=0,1,b,\infty$. Its canonical form is given by \cite{Ronveaux:1995}
\begin{eqnarray}
&& \frac{d^{2}y(z)}{dz^{2}}+\biggl(\frac{\gamma}{z}+\frac{\delta}{z-1}+\frac{\epsilon}{z-b}\biggr)\frac{dy(z)}{dz}\nonumber\\
&& +\frac{\alpha\beta-q}{z(z-1)(z-b)}y(z)=0,
\label{eq:canonical_form_general_Heun}
\end{eqnarray}
where $y(z)=\mbox{HeunG}(b,q;\alpha,\beta,\gamma,\delta;z)$ is the general Heun function, which is simultaneously a local Frobenius solution around two singularities $s_{1}$ and $s_{2}$, where $s_{1},s_{2} \in \{0,1,b\}$. It is analytic in some domain including both these singularities. The parameters $b$, $q$, $\alpha$, $\beta$, $\gamma$, $\delta$, and $\epsilon$ are generally complex and arbitrary, where $b$ is such that $b \neq 0,1$, and they are related by $\gamma+\delta+\epsilon=\alpha+\beta+1$. If $\gamma \neq 0,-1,-2,\ldots$, the general Heun function is analytic in the disk $|z| < 1$, and the following Maclaurin expansion applies \cite{MathAnn.33.161}
\begin{equation}
\mbox{HeunG}(b,q;\alpha,\beta,\gamma,\delta;z)=\sum_{j=0}^{\infty}c_{j}z^{j},
\label{eq:serie_HeunG_todo_z}
\end{equation}
where
\begin{eqnarray}
	b\gamma c_{1}-qc_{0} & = & 0,\nonumber\\
	X_{j}c_{j+1}-(Q_{j}+q)c_{j}+P_{j}c_{j-1} & = & 0 \quad (\mbox{for}\ j \geq 1),
\label{eq:recursion_General_Heun}
\end{eqnarray}
with $c_{0}=1$ and 
\begin{eqnarray}
	P_{j} & = & (j-1+\alpha)(j-1+\beta),\nonumber\\
	Q_{j} & = & j[(j-1+\gamma)(1+b)+b\delta+\epsilon],\nonumber\\
	X_{j} & = & b(j+1)(j+\gamma).
\label{eq:P_Q_X_recursion_General_Heun}
\end{eqnarray}
Thus, these expressions leads to
\begin{equation}
\mbox{HeunG}(b,q;\alpha,\beta,\gamma,\delta;0) \sim 1.
\label{eq:HeunG}
\end{equation}
In addition, the solutions of Eq.~(\ref{eq:canonical_form_general_Heun}) corresponding to the exponents $\alpha$ and $\beta$ at $z=\infty$ are given, respectively, by
\begin{eqnarray}
&& z^{-\alpha}\mbox{HeunG}\biggl(\frac{1}{b},\alpha(\beta-\epsilon)+\frac{\alpha}{b}(\beta-\delta)-\frac{q}{b};\nonumber\\
&& \alpha,\alpha-\gamma+1,\alpha-\beta+1,\delta;\frac{1}{z}\biggr)
\label{eq:alpha_Heun_infinity}
\end{eqnarray}
and
\begin{eqnarray}
&& z^{-\beta}\mbox{HeunG}\biggl(\frac{1}{b},\beta(\alpha-\epsilon)+\frac{\beta}{b}(\alpha-\delta)-\frac{q}{b};\nonumber\\
&& \beta,\beta-\gamma+1,\beta-\alpha+1,\delta;\frac{1}{z}\biggr).
\label{eq:beta_Heun_infinity}
\end{eqnarray}
On the other hand, the general Heun function becomes a polynomial of degree $n$ if it satisfies the so-called $\alpha$-con\-di\-tion, which is given by \cite{Ronveaux:1995}
\begin{equation}
\alpha=-n,
\label{eq:alpha-condition}
\end{equation}
where $n=0,1,2,\ldots$. Such polynomial solutions are denoted by $\mbox{Hp}_{n}(z)=\mbox{HeunG}(b,q;-n,\beta,\gamma,\delta;z)$ and can be written as
\begin{equation}
\mbox{Hp}_{n}(z)=\sum_{\nu=0}^{\infty}c_{\nu}z^{\nu},
\label{eq:polynomial_solutions}
\end{equation}
where the coefficients $c_{\nu}$ are given by
\begin{equation}
-(Q_{0}+q)c_{0}+X_{\nu}c_{1}=0,
\label{eq:cnu}
\end{equation}
\begin{equation}
P_{\nu}c_{\nu-1}-(Q_{\nu}+q)c_{\nu}+X_{\nu}c_{\nu+1}=0,
\label{eq:PQRnu}
\end{equation}
for $\nu=1,2,\ldots,n-1$, where the parameters $P_{\nu}$, $Q_{\nu}$, and $X_{\nu}$ are given by Eq.~(\ref{eq:P_Q_X_recursion_General_Heun}). These equations are consistent if, and only if, the accessory parameter $q$ was chosen properly, which means that there is a polynomial equation of degree $n+1$ for the determination of such a parameter. We will choose the following notation for these eigenvalues: $q_{n;m}$, where $m$ runs from $0$ to $n$. Thus, the corresponding general Heun polynomials are now denoted as $\mbox{Hp}_{n;m}(z)$.

The explicit form of the first three general Heun polynomials can be obtained as follows. For $n=0$, we have
\begin{equation}
\mbox{Hp}_{0;m}(z)=c_{0}=1,
\label{eq:Hp_0,m}
\end{equation}
where the eigenvalues $q_{0;m}$ must obey
\begin{equation}
c_{1}=0,
\label{eq:c_1}
\end{equation}
where
\begin{equation}
-qc_{0}+b \gamma c_{1}=0,
\end{equation}
which implies
\begin{equation}
c_{1}=\frac{q}{b\gamma},
\end{equation}
and then we have that
\begin{equation}
q_{0;0}=0.
\label{eq:q_0,0}
\end{equation}
Thus, the first general Heun polynomial is given by
\begin{equation}
\mbox{Hp}_{0;0}(z)=1.
\label{eq:Hp_0,0}
\end{equation}
Now, for $n=1$, we have
\begin{equation}
\mbox{Hp}_{1;m}(z)=c_{0}+c_{1}z=1+\frac{q_{1;m}}{b\gamma}z,
\label{eq:Hp_1,m}
\end{equation}
where the eigenvalues $q_{1;m}$ must obey
\begin{equation}
c_{2}=0,
\label{eq:c_2}
\end{equation}
where
\begin{equation}
P_{1}c_{0}-(Q_{1}+q)c_{1}+R_{1}c_{2}=0,
\end{equation}
which implies
\begin{equation}
c_{2}=\frac{[\gamma(1+b)+b\delta+\epsilon+q]q-b\alpha\beta\gamma}{2b^{2}\gamma(1+\gamma)},
\end{equation}
and then we have that
\begin{equation}
q_{1;m}=\frac{-[\gamma(1+b)+b\delta+\epsilon] \pm \sqrt{\Delta}}{2},
\label{eq:q_1,m}
\end{equation}
where $\Delta=[\gamma(1+b)+b\delta+\epsilon]^{2}+4b\alpha\beta\gamma$. Note that the signs $-$ and $+$ stand for $m=0$ and $m=1$, respectively. Thus, the second and third general Heun polynomials are given, respectively, by
\begin{equation}
\mbox{Hp}_{1;0}(z)=1+\frac{-[\gamma(1+b)+b\delta+\epsilon] - \sqrt{\Delta}}{2b\gamma}z
\label{eq:Hp_1,0}
\end{equation}
and
\begin{equation}
\mbox{Hp}_{1;1}(z)=1+\frac{-[\gamma(1+b)+b\delta+\epsilon] + \sqrt{\Delta}}{2b\gamma}z.
\label{eq:Hp_1,1}
\end{equation}
Finally, the corresponding Hamiltonian $\mathcal{H}$ in classical mechanics is given by
\begin{eqnarray}
\mathcal{H}(q,p,t) & = & \frac{-1}{t(t-1)}\{q(q-1)(q-t)p^{2}\nonumber\\
&& +[\gamma (q-1)(q-t)+\delta q(q-t)\nonumber\\
&& +\epsilon q(q-1)]p+\alpha\beta q\}.
\label{eq:Hamiltonian_general_Heun}
\end{eqnarray}
From now on, $q$ and $p$ are the canonical coordinate and momentum, respectively, and $t$ is the scaling parameter (which can be considered as time). Then, if $q$ and $p$ are quantum observables, we can write $\mathcal{H}(q,p,t) y=\lambda y$, where $\lambda$ is the eigenvalue (which can be considered as energy).
%
%
\subsection{The deformed Heun equation}
Next, let us talk about the deformed Heun equation. In fact, Slavyanov and Lay \cite{Slavyanov:2000} presented the Heun class of equations, which includes the confluent cases, in an extended form by adding an apparent singularity to each equation. However, none of these forms fits our case. Then, we will follow their ideas in order to discuss a particular case of the deformed Heun equation, which can be written as
\begin{eqnarray}
&& \frac{d^{2}y(z)}{dz^{2}}+\biggl(\frac{\gamma}{z}+\frac{\delta}{z-1}+\frac{\epsilon}{z-t}-\frac{1}{z-q}\biggr)\frac{dy(z)}{dz}\nonumber\\
&& +\biggl[\frac{\alpha\beta}{(z-1)(z-t)}+\frac{ht(t-1)}{z(z-1)(z-t)}\nonumber\\
&& +\frac{pq(q-1)}{z(z-1)(z-q)}\biggr]y(z)=0.
\label{eq:canonical_deformed_general_Heun}
\end{eqnarray}
Note that it differs from the (general) Heun equation (\ref{eq:canonical_form_general_Heun}) by two additional terms proportional to $z-q$, which is a simple pole at $z=q$ and plays the role of an apparent singularity. The parameters $h$ and $p$ obey to the following relations
\begin{equation}
h=\mbox{Res}_{z=t}\frac{ht(t-1)}{z(z-1)(z-t)},
\label{eq:h_deformed_Heun}
\end{equation}
\begin{equation}
p=\mbox{Res}_{z=q}\frac{pq(q-1)}{z(z-1)(z-q)}.
\label{eq:p_deformed_Heun}
\end{equation}
Now, an analytical solution of the deformed Heun equation given by Eq.~(\ref{eq:canonical_deformed_general_Heun}) can be written as a Taylor expansion, namely,
\begin{equation}
y(z)=\sum_{k=0}^{\infty}g_{k}(z-q)^{k},
\label{eq:Taylor_expansion}
\end{equation}
where $g_{0}=1$. For simplicity, we will rewrite Eq.~(\ref{eq:canonical_deformed_general_Heun}) as
\begin{equation}
\frac{d^{2}y(z)}{dz^{2}}+P(z)\frac{dy(z)}{dz}Q(z)y(z)=0,
\label{eq:canonical_deformed_general_Heun_2}
\end{equation}
where the coefficients $P(z)$ and $Q(z)$ are given in term of the following Laurent expansions
\begin{equation}
P(z)=\sum_{k=-1}^{\infty}p_{k}(z-q)^{k}=p_{0}+\frac{p_{-1}}{z-q},
\label{eq:P_Laurent_expansion}
\end{equation}
\begin{equation}
Q(z)=\sum_{k=-1}^{\infty}q_{k}(z-q)^{k}=q_{0}+\frac{q_{-1}}{z-q}.
\label{eq:Q_Laurent_expansion}
\end{equation}
Thus, substituting Eqs.~(\ref{eq:Taylor_expansion})-(\ref{eq:Q_Laurent_expansion}) into Eq.~(\ref{eq:canonical_deformed_general_Heun}), we obtain
\begin{eqnarray}
&& g_{k}k(k-1)(z-q)^{k-2}+p_{-1}g_{k}k(z-q)^{k-2}\nonumber\\
&& +p_{0}g_{k}k(z-q)^{k-1}+q_{-1}g_{k}(z-q)^{k-1}\nonumber\\
&& +q_{0}g_{k}(z-q)^{k}=0,
\label{eq:recurrence_1}
\end{eqnarray}
from which we arrive at the following recurrence relation
\begin{equation}
k(k+2)g_{k+2}+[p_{0}(k+1)+q_{-1}]g_{k+1}+q_{0}g_{k}=0,
\label{eq:recurrence_2}
\end{equation}
where $k \geq 0$. From this recurrence relation, we obtain
\begin{eqnarray}
p_{-1}g_{1}+q_{-1}g_{0} & = & 0,\nonumber\\
2(1+p_{-1})g_{2}+(p_{0}+q_{-1})g_{1}+q_{0} & = & 0.
\label{eq:recurrence_3}
\end{eqnarray}
On the other hand, the values for the coefficients $p_{k}$ and $q_{k}$ are directly obtained from Eq.~(\ref{eq:canonical_deformed_general_Heun}). They are given by
\begin{eqnarray}
p_{-1} & = & -1,\\
p_{0}  & = & \frac{\gamma}{q}+\frac{\delta}{q-1}+\frac{\epsilon}{q-t},\\
q_{-1} & = & p,\\
q_{0}  & = & \frac{\alpha\beta}{(q-1)(q-t)}+\frac{ht(t-1)}{q(q-1)(q-t)}\nonumber\\
&& -p\biggl(\frac{1}{q}-\frac{1}{q-1}\biggr).
\label{eq:pk_qk}
\end{eqnarray}
In these terms, Eqs.~(\ref{eq:recurrence_3}) take the following form
\begin{eqnarray}
0 & = & -g_{1}+p,\nonumber\\
0 & = & \biggl(\frac{\gamma}{q}+\frac{\delta}{q-1}+\frac{\epsilon}{q-t}+p\biggr)p+\frac{\alpha\beta}{(q-1)(q-t)}\nonumber\\
&& +\frac{ht(t-1)}{q(q-1)(q-t)}-p\biggl(\frac{1}{q}-\frac{1}{q-1}\biggr).
\label{eq:recurrence_4}
\end{eqnarray}
Therefore, the necessary condition for the point $z=q$ to be an apparent singularity is
\begin{eqnarray}
h(q,p,t) & = & \frac{-1}{t(t-1)}\{q(q-1)(q-t)p^{2}\nonumber\\
&& +[(\gamma-1) (q-1)(q-t)\nonumber\\
&& +(\delta-1) q(q-t)+\epsilon q(q-1)]p\nonumber\\
&& +\alpha\beta q\}.
\label{eq:necessary_condition}
\end{eqnarray}
It is easy to see that this function $h(q,p,t)$ coincides with the classical Hamiltonian given by Eq.~(\ref{eq:Hamiltonian_general_Heun}), concerning to the general Heun equation. However, it differs by an unitary shifting in the parameters $\gamma$ and $\delta$. It can be written as
\begin{eqnarray}
h(q,p,t;\alpha,\beta,\gamma,\delta,\epsilon)=\mathcal{H}(q,p,t;\alpha,\beta,\gamma-1,\delta-1,\epsilon).\nonumber\\
\label{eq:Hamiltonian_shifting}
\end{eqnarray}
The Slavyanov and Lay's explanation for this shifting is that the parameters $\alpha$, $\beta$, $\gamma$, $\delta$, and $\epsilon$ satisfy different Fuchs conditions related to the different cases (general and deformed) of the Heun equations. In the case of a deformed Heun equation, the condition is $\gamma+\delta+\epsilon$ $=$ $\alpha+\beta+3$.
%
%

%
%
\end{document}